# Investigation of the doping effects of Sr-Ta on the Ionic Conductivity of Garnet $Li_7La_3Zr_2O_{12}$ Solid Electrolyte

**Muktai Aote[a], A. V. Deshpande[a,*], Kajal Parchake[a], Anuj Khapekar[a]**

[a]*Department of Physics, Visvesvaraya National Institute of Technology, South Ambazari Road, Nagpur, 440010, Maharashtra, India*

[*]**Corresponding Author:**

**Dr. (Mrs.) A. V. Deshpande**

Department of Physics,

Visvesvaraya National Institute of Technology,

South Ambazari Road, Nagpur

Maharashtra, 440010 (India)

E-mail Address: avdeshpande@phy.vnit.ac.in

Ph.No.: 0-712-280-1173

## Abstract

A solid electrolyte having the ionic conductivity comparable to that of conventional liquid electrolyte can be used in All Solid State Batteries (ASSB's). The series $Li_{6.75+x}La_{3-x}Sr_xZr_{1.75}Ta_{0.25}O_{12}$ (x = 0 to 0.20) was synthesized to improve the ionic conductivity of garnet $Li_7La_3Zr_2O_{12}$ (LLZO). The structural, physical and morphological investigations have been carried out for all the synthesized samples using X ray diffraction, density measurement and scanning electron microscopy respectively. The results of electrochemical analysis showed that the maximum room temperature ionic conductivity of 3.5 x $10^{-4}$ S/cm and minimum activation energy of 0.29 eV is achieved by the 0.05 Sr ceramic sample sintered at 1050º C. The DC conductivity measurement confirmed the dominance of ionic conduction in the prepared ceramic samples. The highest ionic conductivity with the minimum activation energy makes the 0.05 Sr ceramic sample a suitable choice as solid electrolyte for All Solid State Lithium Ion Batteries (ASSLIB's).

## 1. Introduction

Lithium ion batteries (LIB's) play an important and major role in the advancement and rapid growth of battery industries. All this while, the conventional Li rich liquid electrolytes have been used in LIB's. However, the main drawback related to conventional liquid electrolytes, which have been extensively used in LIB's, is growth of dendrite after repetitive duty cycles [1]. Moreover, the leakage of such liquid electrolyte in electric gadgets has now become the serious issue. All these reasons are responsible to the increase in incidents of flammability which raised the concern regarding consumer's safety [2]. Therefore, in order to overcome these shortcomings of liquid electrolytes, the research has been directed towards the solid electrolytes. In order to be used as an electrolyte in all solid state lithium ion batteries (ASSLIB's), solid electrolyte needs to possess the properties as [1–3]:

1) Room temperature Li ion conductivity in the range of $10^{-3}$- $10^{-4}$ S/cm
2) Mechanical and chemical stability
3) High energy and power density
4) Wide electrochemical potential window
5) Long duty cycle with high specific capacity

There are various categories of solid electrolytes that have been investigated till now [4]. Out of them, the most studied solid electrolytes are discussed here. One of its types is sulfide electrolytes [5]. This solid electrolyte has the high room temperature Li ion conductivity with large potential window comparable to liquid electrolytes. Also, it showed great deformability as it is soft in nature [6]. The high density solid electrolyte which eases the Li-ion conduction can be synthesized using metal sulfides. But the main drawbacks of these sulfide solid electrolytes are their instability in ambient atmosphere and extreme reactivity towards water which make it difficult to synthesize on large scale application [7]. Along with sulfide electrolytes, polymer solid electrolytes were widely explored due to their mechanical strength along with superior elastic behavior. This helped in minimizing the interfacial resistance. However, it failed to achieve the desired range of Li-ion conductivity as well as the optimum electrochemical potential window required for practical application in ASSLIB's [8–10]. The solid electrolyte in the form of thin film has also been studied. But its critical synthesis procedure and low ionic conductivity became the constraints in its way of practical use as an electrolyte [1,11–13].

Along with the above mentioned solid electrolytes, one more type is inorganic solid electrolyte which has been extensively studied and the research is still going on with it. It is observed that, in the conduction of Li-ion, the crystal structure plays an essential role [2]. Some of the inorganic solid electrolytes with different lattice arrangements are listed here. $Li_{1.4}Al_{0.4}Ti_{1.6}(PO_4)_3$ with NASICON crystal structure [14,15], $Li_{0.35}La_{0.55}TiO_3$ having Perovskite crystal structure [1,16,17], $Li_2S$-$SiS_2$-$Al_2S_3$ representing the thio- LISICON crystal structure [2,18] and $Li_7La_3Zr_2O_{12}$(LLZO) having garnet type crystal structure [1,2,19,20] have been thoroughly investigated.

Except garnet structured LLZO, the other above mentioned solid electrolytes, have one or the other disadvantages like low ionic conductivity, thermal and chemical instability, narrow potential window associated with them [1,20]. Thus, more focus has been driven towards the detailed study of garnet structured oxide LLZO solid electrolyte. Garnet LLZO gained the researchers attention due to its high room temperature Li ion conductivity, stability against the Li metal anode, superior mechanical strength and comparable range of operating potential window which make it a promising candidate as solid electrolyte in practical application in ASSLIB's [2,19]. Normally, LLZO has tetragonal phase ($I4_1/acd$) which gives the Li-ion conductivity in the range of $10^{-6}/10^{-7}$ S/cm due to its ordered and thermodynamically stable structure [1]. On the other hand, after sufficient heat treatment this tetragonal phase of LLZO can be transformed into the cubic phase (Ia-3d), which has the high Li-ion conductivity by two orders of magnitude than initial phase at room temperature. However, due to the disordered behavior of lattice element, cubic phase is unstable at room temperature [20]. The review of literature on LLZO revealed that, in stabilizing the highly conducting cubic phase of garnet LLZO, the Li-ion vacancies play an important role which can be created by the doping of supervalent cations in the lattice of LLZO [1,2,20].

In the crystal lattice of garnet LLZO, the supervalent doping can be done either at the single site or at the dual sites respectively. Some of the studies show the increment in Li-ion conductivity by simultaneous substitution at all the sites i.e. Li, La and Zr in LLZO [2]. The substitution at Li site affects the Li-ion conductivity by changing the regular Li-ion distribution whereas the substitution at Zr site can create the Li-ion vacancies which eventually help in the stabilization of cubic phase [21]. Moreover, the substitution at La site helps in the regulation of Li-ion migration channel by increasing the Li content [22]. Till now various elements like Al, Ge, Ta, Ga, Sr, Ca, Ba,Ti, Mn, Zn, Nb, W, Fe have been substituted in garnet LLZO by opting the single doping as well as dual doping strategies [23–36]. The reviewed studies supported the fact that, the substitution of Ta at the Zr site undoubtedly helps to stabilize the conducting cubic phase by creating the Li-ion vacancies and also helps in the increment in ionic conductivity [1,37,38].However, to transform the tetragonal phase to cubic phase, high sintering temperature is required which eventually results into Li loss, affecting the total ionic conductivity of the material [2]. Thus, in order to get cubic phase at relatively lower sintering temperature, some sintering additives were used. Out of them, Sr recently emerged as a sintering additive which also helps in the densification of the structure[27,39]. Moreover, substitution of Sr at lanthanum site can lead to increase in total Li content which results into high ionic conductivity [2].

Thus, in this work Sr and Ta has been selected as supervalent dopants to be doped in the garnet LLZO at La and Zr sites respectively. Sr and Ta were substituted as a sintering aid and cubic phase stabilizing agent respectively. The content of Ta was kept constant at 0.25 atoms per formula unit (a.p.f.u.) [1,2]. While, the content of Sr has been varied from 0 to 0.20 a.p.f.u. The synthesized samples are represented according to the content of Sr as 0 Sr, 0.05 Sr, 0.10 Sr, 0.15

Sr and 0.20 Sr respectively. All the synthesized samples were investigated with various structural and electrochemical characterizations and the results are discussed below.

## 2. Experimental Work

### *2.1. Material Preparation*

By varying the Sr content with fixed Ta, the series $Li_{6.75+x}La_{3-x}Sr_xZr_{1.75}Ta_{0.25}O_{12}$ (x = 0 to 0.20) was prepared using conventional solid state reaction method. For this, all the required precursors namely, $Li_2CO_3$ (Merck, >99.9%), $La_2O_3$, $ZrO_2$, $SrCO_3$and $Ta_2O_5$ (Sigma Aldrich, >99.99%) were stoichiometrically measured and then hand mixed in an agate mortar followed by wet mixing using acetone as an aqueous medium. During this process, the excess 10% $Li_2CO_3$was added in order to minimize the Li loss during heat treatment. The prepared powder was then transferred into alumina crucible and kept into the muffle furnace for calcination at 900ºC for 8h. After the calcination process, the calcined powder was allowed to cool down till ambient temperature and again crushed into a fine powder. The diaset of around 10 mm diameter and 1.5 mm thickness was used to make pellets using hydraulic press with the pressure of 4 tons. The formed pellets were kept in the bed of mother powder and the crucible was covered with alumina lid to prevent contamination during sintering at 1050ºC for 8h. The sintered pellets were then subjected to various characterizations mentioned as follows.

### *2.2.Materials Characterizations*

Various characterizations like XRD, SEM, Density measurement, and Electrochemical Impedance spectroscopy were done for the sintered pellets. The finely crushed powder of sintered pellet was used for the phase identification using X-ray diffraction (RIGAKU diffractometer). Here Cu-kα (wavelength = 1.54 Å) was used as a radiation source, keeping step size of 0.02º with the scan rate of 2º/min. The XRD data was collected in the range of 10º to 70º. The Archimedes' principle was used to calculate the densities of the samples with the help of K-15 Classic (K-Roy) instrument. Here, toluene was used as an immersion medium. The structural morphology and elemental analyses were studied using the scanning electron microscopy having the attachment of energy dispersive spectroscope with it (JSM-7600 F/JEOL). For the impedance and activation energy measurement, NOVOCONTROL impedance analyzer was used within the frequency range of 20 Hz to 20 MHz and the temperature varied from room temperature to 150ºC. The pellets were coated with silver paste from both the sides. It maintained the ohmic contact with the silver electrodes which acts as ion blocking electrodes. To confirm the ionic conduction within the sample, the DC Conductivity measurement was done using KEITHLEY 6512 programmable electrometer.

## 3. Results and Discussion

### *3.1. X Ray Diffraction Study*

The X ray diffraction patterns of all the prepared samples of the series $Li_{6.75+x}La_{3-x}Sr_xZr_{1.75}Ta_{0.25}O_{12}$ with x ranging from 0 to 0.20 are presented in Fig. 1(a). From the figure, it can be clearly observed that, all the synthesized ceramic samples possessed the conducting cubic phase which has been exactly matched with the garnet cubic structure of $Li_5La_3Nb_2O_{12}$ (JCPDS File No. 45.0109. All the peaks are indicated with their respective (h, k, l) values. It can be seen that, there is no impurity peak present within the XRD patterns of $Li_{6.75+x}La_{3-x}Sr_xZr_{1.75}Ta_{0.25}O_{12}$. In contrast to this, the study reported by Changwei Lin et.al. [40] showed that, for the minimum Sr content of 0.05 and 0.10 a.p.f.u. the impurity phase was observed which was attributed to the Li loss. Thus, the present study confirmed that, the optimum amount of Ta helped the Sr doped LLZO structure to stabilize the cubic phase at relatively lower sintering temperature without any external impurity peaks [1,2]. Also, with the insertion of Sr, there is no change in the phase which suggests that, the doping of Sr at La site did not affect the phase of garnet LLZO [27]. Moreover, from the Fig.1 (a), it can be observed that, with the increase in content of Sr, the peak intensity and sharpness have been increased. This can be attributed to the sintering ability of Sr which increased the crystallinity of the peak [41,42]. The successful insertion of Sr in the lattice of LLZO can be depicted from the Fig. 1(b) where the shifting of peaks can be observed. With the increase in Sr content, the peak in the range of 16º- 17.5º shifted towards the lower theta value. This may be due to the difference in ionic radius of Sr (1.26Å) and La (1.16Å) which might expand the lattice [27]. However, as the Sr content exceeded the optimum content of 0.15 a.p.f.u., the peak again shifted towards the higher angle. This confirmed the fact that, beyond the optimum limit of doping, the excess amount of Sr could not enter the crystal lattice of LLZO and might be present at the grain boundaries of the ceramic sample [41].

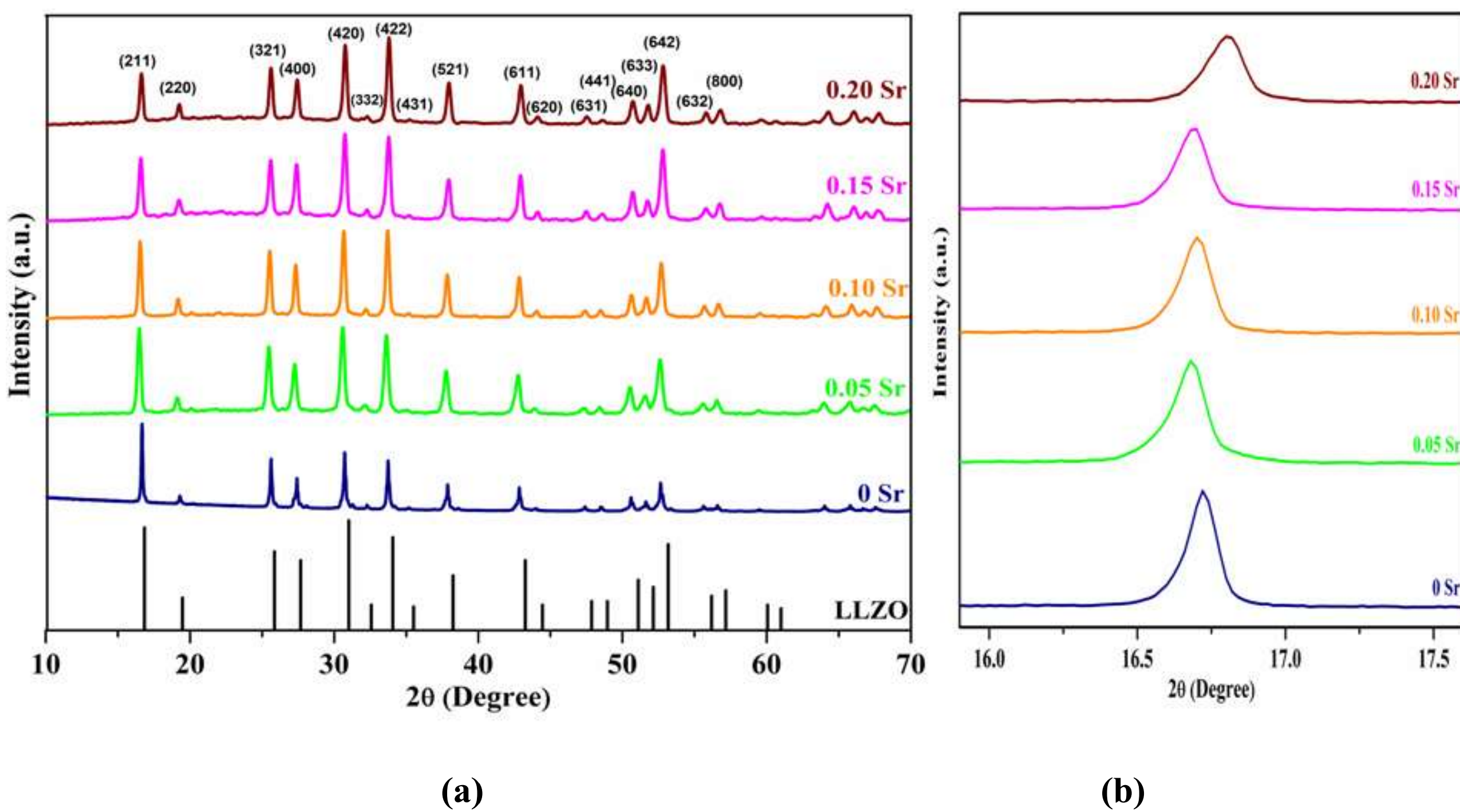

**Fig. 1: (a) XRD patterns and (b) Shifting of (211) peak of $Li_{6.75+x}La_{3-x}Sr_xZr_{1.75}Ta_{0.25}O_{12}$ with x = 0 to 0.20.**

### *3.2.Density Study*

The density plays a crucial role in governing the ionic conductivity, as the sample with highest density minimizes the Li ion migration pathways [2]. The densities of all the synthesized ceramic samples of the series $Li_{6.75+x}La_{3-x}Sr_xZr_{1.75}Ta_{0.25}O_{12}$ (x = 0 to 0.20) were measured using Archimedes' principle. The variation in density and relative density is shown in Fig.2. From the figure, it can be observed that with the increase in Sr content, the density also increased. The maximum density of 4.65 g/cm$^3$ was measured for the 0.05 Sr with the highest relative density of 93%. The obtained result is in good agreement with the previous study by Xiaoren Zhou et.al. [43], which reported the highest density with the 0.05 a.p.f.u. Sr in Mo doped LLZO. Even though there is a slight decrease in the density of the sample beyond 0.05 Sr content, the relative density is more than 90%. This can be attributed to the sintering ability of Sr which helped in the densification of the crystal structure [39,41,44,45]. Also the decrease in density beyond optimum content of 0.05 a.p.f.u., can be correlated with the presence of voids within the structure, which are formed due to the higher ionic radius of Sr that affects the process of densification and can be observed from the SEM micrographs study.

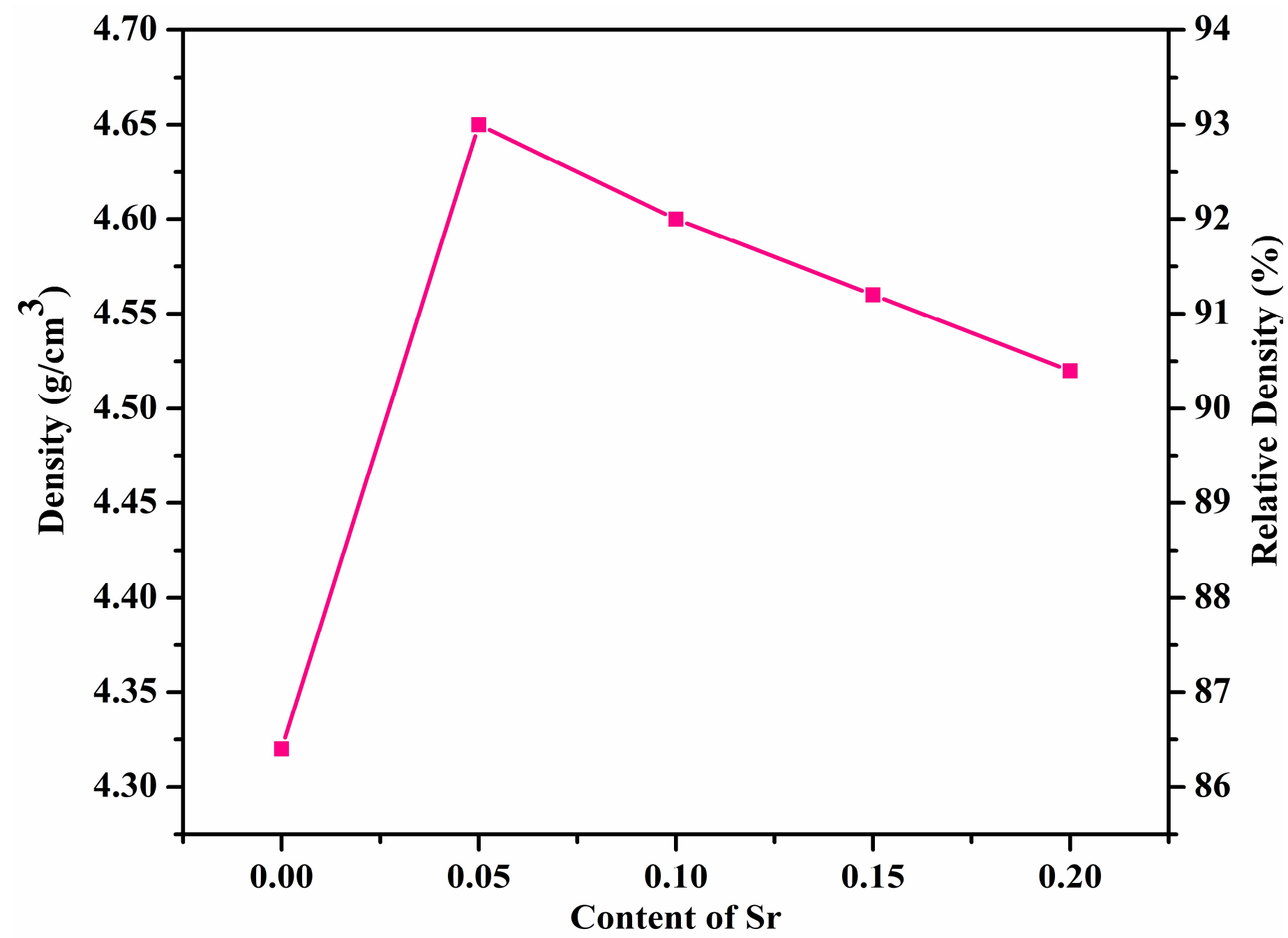

**Fig.2: Variation in Density and Relative density with the content of Sr in**

**$Li_{6.75+x}La_{3-x}Sr_xZr_{1.75}Ta_{0.25}O_{12}$.**

### *3.3.Surface Morphological Study*

The surface micrographs of all the synthesized samples with Sr content varying from 0 to 0.20 are given in Fig. 3(a-e). Fig. 3 (a) shows the morphology of 0 Sr ceramic sample which has visible pores as well as non-uniform grain structure. But with the increase in Sr content, the uniform grain growth can be seen in Fig.3 (b-e). Among all the ceramic samples, the 0.05 Sr ceramic sample showed the compact and dense microstructure without any voids as indicated in Fig. 3(b). All the grains are well connected with the neighboring grains, avoiding the possibility of dendrite growth which affects the ionic conductivity. This result is in accordance with the highest density value obtained for 0.05 Sr ceramic sample as depicted in Fig.2. This can be attributed to the sintering ability of Sr which directly helped in the densification of the crystal structure [27,40]. This result confirmed the fact that, the doping content plays an important role to achieve the compact structure with high density and minimum grain boundaries. However, with the further increase in Sr content, the grain size increases but at the same time there is formation of pores within the structure. This may be due to the volatilization of Sr at the higher sintering temperature which also reduces the relative density of the synthesized samples [41]. The presence of pores eventually affects the total Li ion conductivity of the sample [1,2]. Here, Fig.4 shows the cross sectional image of 0.05 Sr ceramic sample with the histogram of average particle size distribution. The magnified image supports the claim of compact structure with suitable grain growth. The average particle size for 0.05 Sr ceramic sample was found to be 3.3±0.03μm which is well supported by the result obtained by Tianxiang Ning et.al.[27].

Fig. 5 shows the elemental mapping with EDX spectra of the 0.05 Sr ceramic sample. From the images, it can be clearly observed that, all the constituent elements i.e. La, Zr, Ta and Sr are present within the material with uniform distribution over the surface. The presence of Ta around the grain boundary as well as the successful insertion of Sr in lattice of LLZO was also confirmed by the shifting of XRD peak in Fig.1(b) [1].

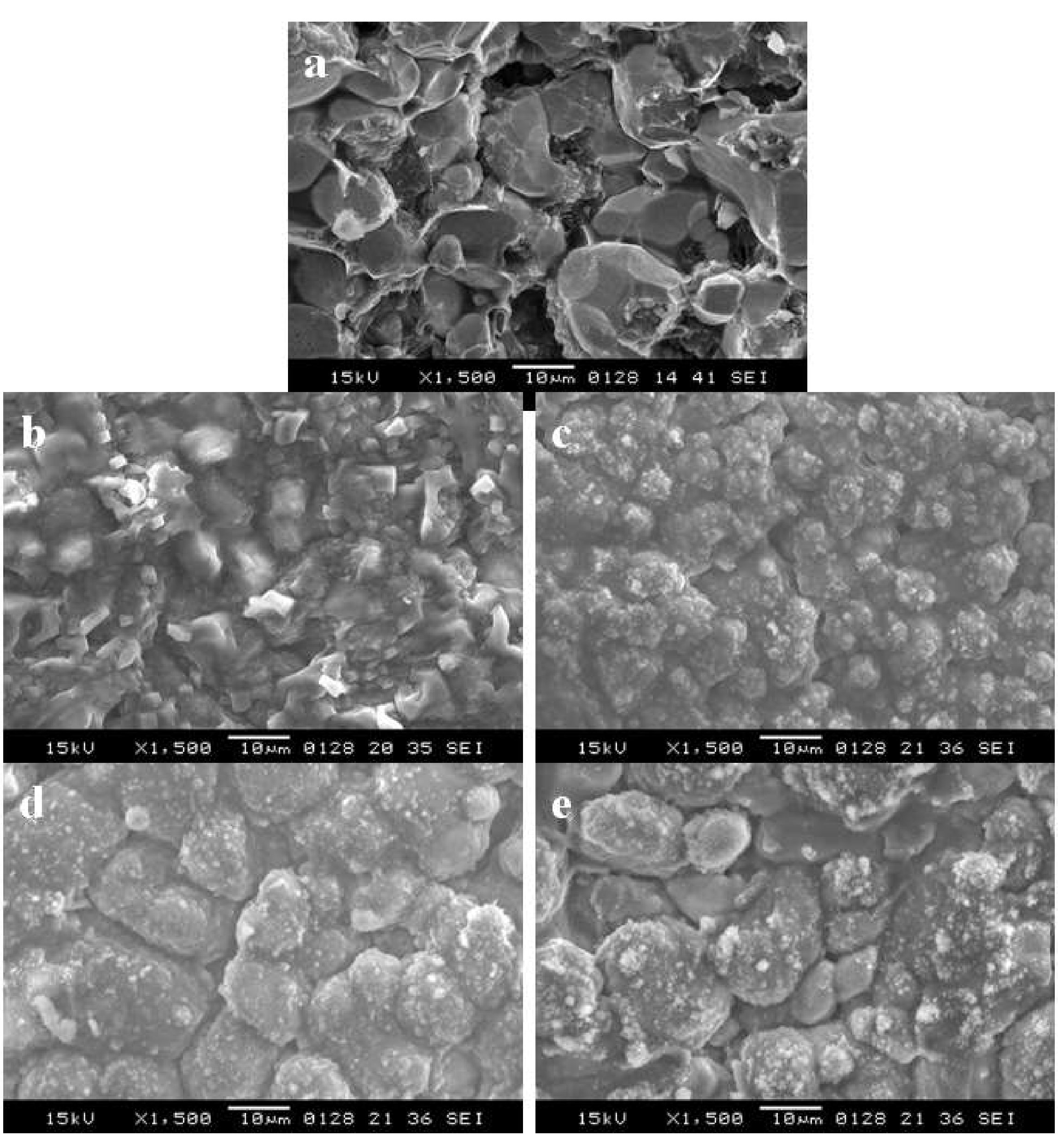

**Fig. 3: Surface morphology images of the ceramic samples of the series $Li_{6.75+x}La_{3-x}Sr_xZr_{1.75}Ta_{0.25}O_{12}$ with x = a) 0, b) 0.05, c) 0.10, d) 0.15, e) 0.20 respectively.**

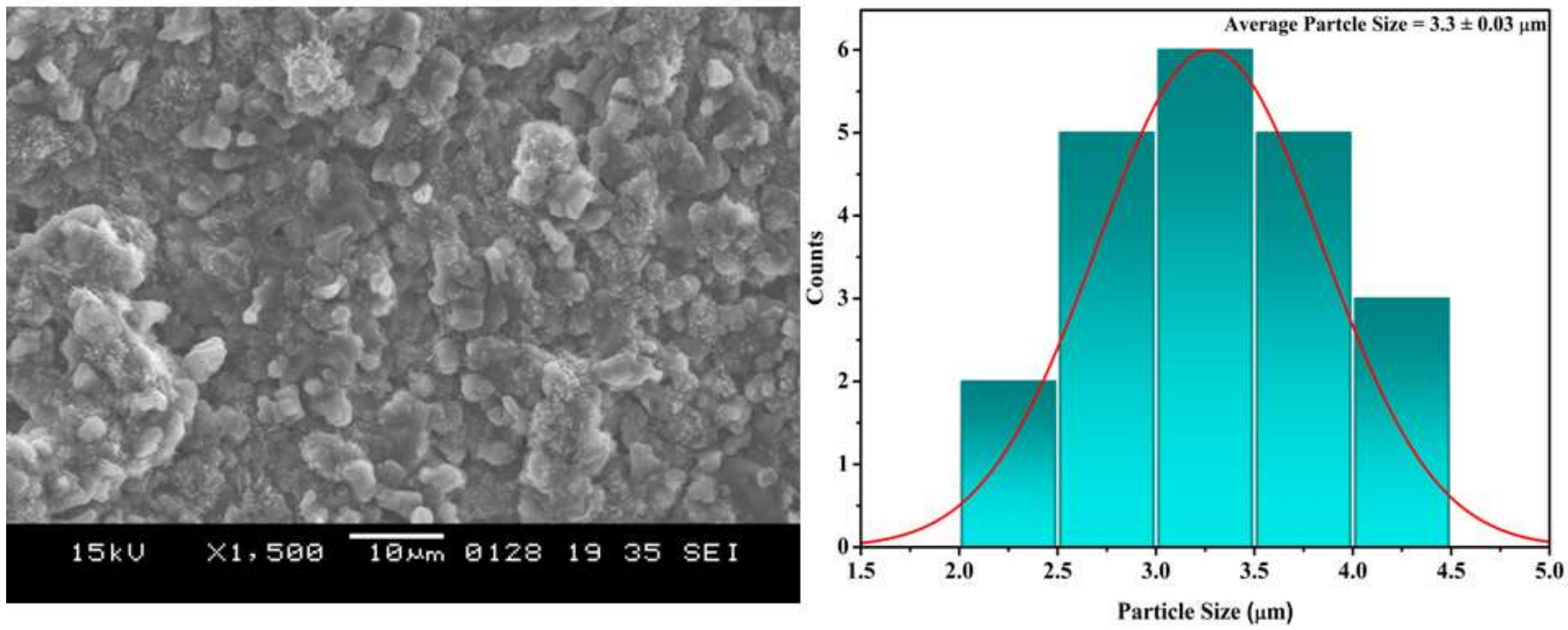

**Fig. 4: Cross sectional surface micrograph with average particle size distribution of 0.05 Sr ceramic sample.**

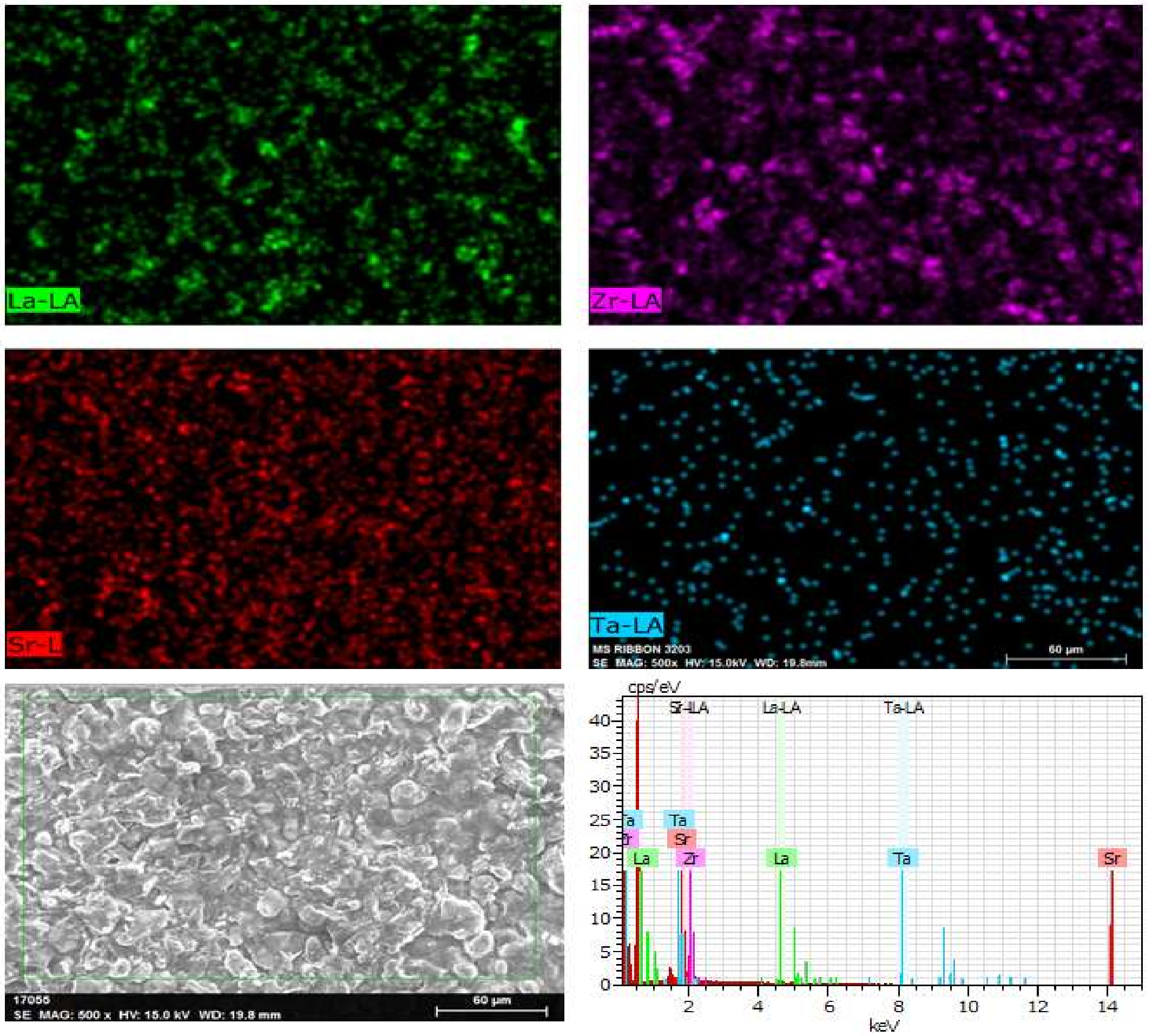

**Fig. 5: Elemental mapping with EDX spectra of 0.05 Sr ceramic sample.**

### *3.4. AC conductivity Study*

#### *3.4.1. Impedance Plots*

Fig. 6 (a) displays the impedance plots of all the synthesized samples of $Li_{6.75+x}La_{3-x}Sr_xZr_{1.75}Ta_{0.25}O_{12}$ with x varying from 0 to 0.20 at room temperature. The nature of graph consists of a semicircle with the extended tail. The intercept made by semicircle on the real axis (Zs') in higher frequency region, gives the value of impedance of each synthesized ceramic sample. Whereas, the appearance of tail in lower frequency region was associated with the ion blocking nature of Ag electrodes. Such pattern of nyquist plot primarily confirms the ionic conduction within the sample [1,2].The formula, $\sigma_{total} = t/RA$ was used to calculate the ionic conductivity of the synthesized ceramic samples where, $\sigma_{total}, R$ ,and $A$ represents the ionic conductivity, thickness, resistance and the area of the sample respectively. It can be clearly observed that, the inset graph of 0.05 Sr in Fig. 6 (a), has the minimum impedance whereas the sample with 0 Sr possessed the highest impedance among all the synthesized samples in $Li_{6.75+x}La_{3-x}Sr_xZr_{1.75}Ta_{0.25}O_{12}$ (x = 0-0.20). The circuit fitting was done as shown in Fig. 6(b). The minimum impedance offered by 0.05 Sr sample gives the highest room temperature ionic conductivity of 3.5 x $10^{-4}$ S/cm. Though the obtained conductivity is slightly lower than the earlier reported study, but the sintering time and temperature as well as the content of both Sr and Ta made the obtained conductivity distinguishable from the previous result [27]. Moreover, the obtained conductivity is greater than the Nb and Sr doped LLZO where 0.05 Sr content gave the minimum conductivity with the inclusion of tetragonal phase [40]. Thus, the highest conductivity can be attributed to the introduction of Ta and Sr in garnet LLZO which caused the synergistic effect not only by stabilizing the cubic phase but also by lowering the sintering temperature which eventually enhanced the grain growth and simultaneously helped in Li ion migration. Also, the ionic conductivity value is in well agreement with the earlier discussed results of highest density and dense microstructure. But as the Sr content exceeds the optimum limit of 0.05 a.p.f.u., there is a decrease in ionic conductivity. This can be attributed to the lower density as well as the open pores within the structure which obstructed the Li ion migration pathways and lead to decrease in the total ionic conductivity. This was also confirmed by the XRD peak shifting, which suggested that the excess of Sr could not enter the LLZO lattice and hence affected the ionic conductivity. Hence, the result confirmed that, the minimum content of Sr (0.05) with Ta helped in grain growth with reducing the grain boundary resistance and enhancing the ionic conductivity in garnet LLZO.

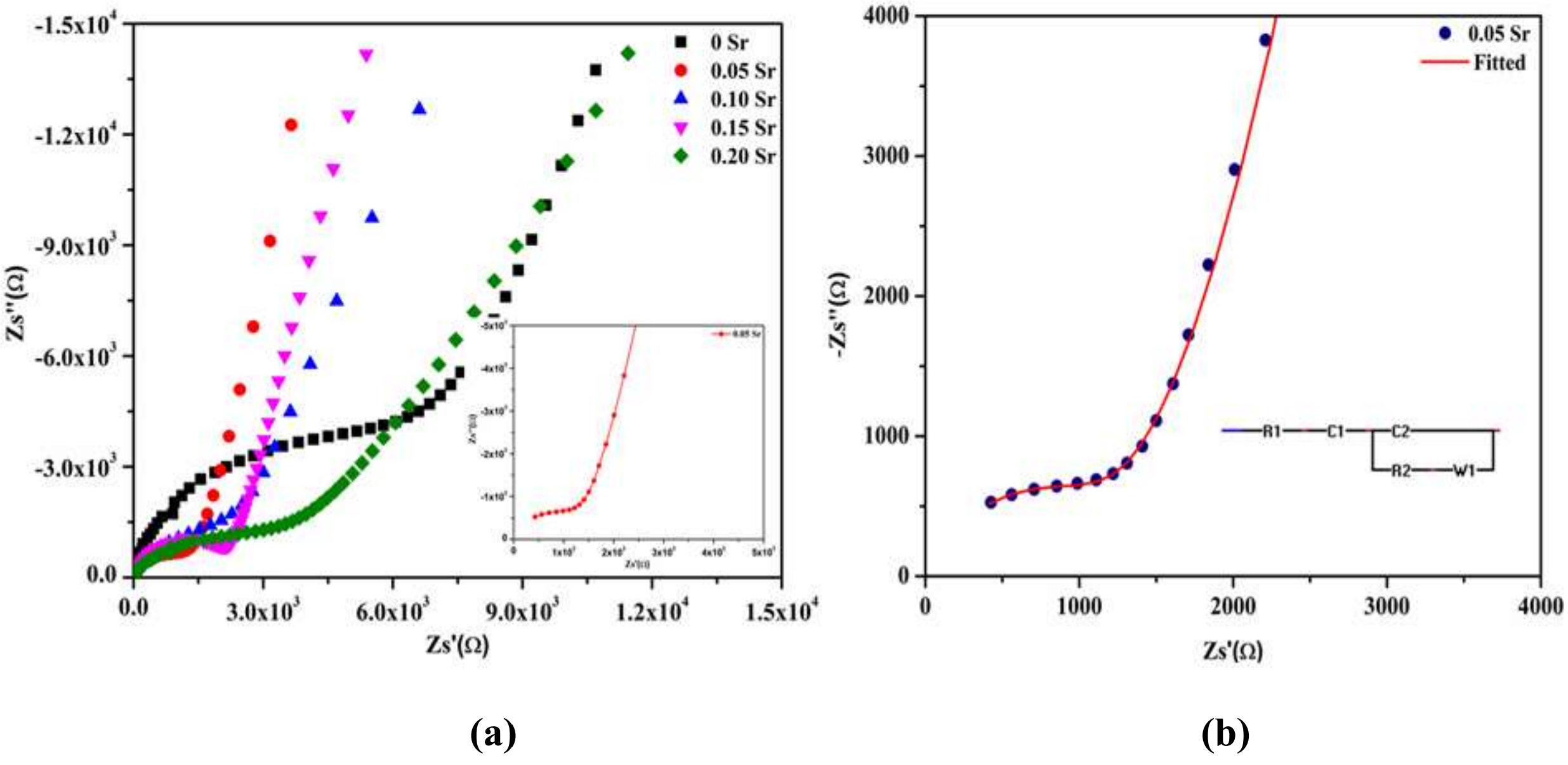

**Fig.6: (a) Nyquist plots of $Li_{6.75+x}La_{3-x}Sr_xZr_{1.75}Ta_{0.25}O_{12}$ with x varying from 0 to 0.20 at room temperature.**

**(b) Fitted Nyquist plot of 0.05 Sr ceramic sample.**

### *3.4.2. Arrhenius Plots*

The activation energies associated with total ionic conductvity of all the synthesized ceramic samples in series $Li_{6.75+x}La_{3-x}Sr_xZr_{1.75}Ta_{0.25}O_{12}$ (x = 0 – 0.20) were calculated using the Arrhenius equation. The respective Arrhenius plots are represented in Fig. 7(a). The data in the temperature range of 25° C to 150° C was utilized to calculate the activation energy using the Arrhenius equation,$\sigma(T) = \sigma_0 \exp\left(-E_a/K_B T\right)$. Here, $\sigma$ is the ionic conductivity, $\sigma_0$ is the pre-exponential factor, $E_a$ is the activation energy, $K_B$ is the Boltzmann constant and $T$ is the temperature (k) respectively. From the Fig. 7(a), it can be clearly observed that, the minimum activation energy is exhibited by the 0.05 Sr ceramic sample which also possessed the highest room temperature total ionic conductivity. The lowest activation energy of 0.29 eV is associated with the highest room temperature ionic conductivity of 3.5 x $10^{-4}$ S/Cm. The respective activation energies and ionic conductivity values of all the synthesized samples are mentioned in table 1. Also, the variation in activation energy with the ionic conductivity for all the samples can be seen in Fig. 7(b). The lower activation energy of 0.05 Sr ceramic sample can be attributed to the ability of Sr as a sintering aid which improved the bottleneck migration pathways of the Li ions by minimizing the possible lattice distortion[27,40,41]. Moreover, the addition of 0.05 Sr helped to maintain the optimum content of Li which yields the highest ionic conductivity and lower activation energy [1,2]. Thus, it can be confirmed that, the minimum content of 0.05 a.p.f.u. of Sr along with Ta, is useful for achieving the highest ionic conductivity with lower activation energy in garnet LLZO.

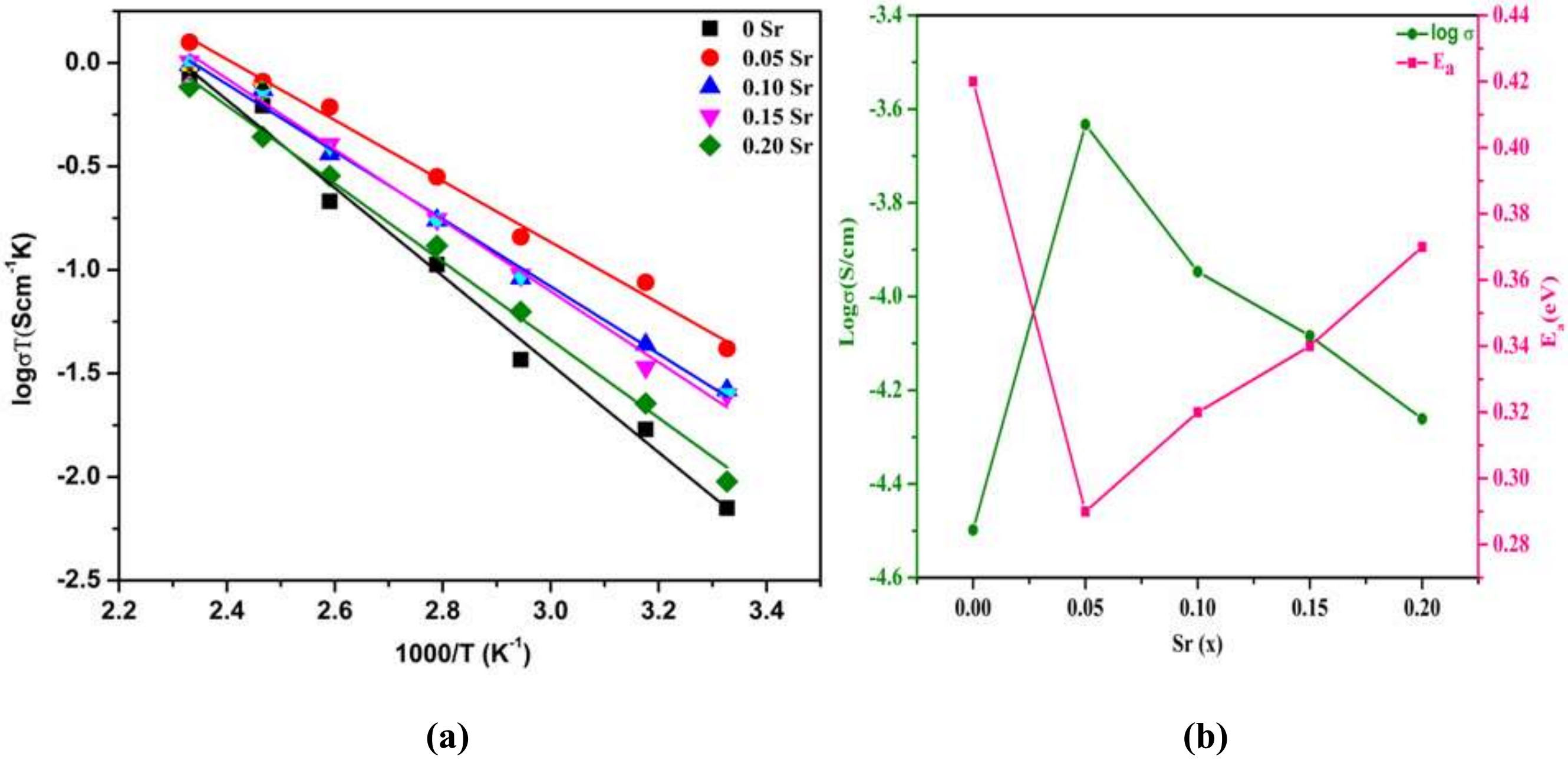

**Fig.7: (a) Arrhenius plots and (b) Variation of ionic conductivity and activation energy of series $Li_{6.75+x}La_{3-x}Sr_xZr_{1.75}Ta_{0.25}O_{12}$ with x varying from 0 to 0.20.**

**Table 1: Values of Ionic conductivity and activation energy of $Li_{6.75+x}La_{3-x}Sr_xZr_{1.75}Ta_{0.25}O_{12}$ (x = 0 to 0.20).**

| Content of Sr (x) | Ionic conductivity (S/cm) | Activation Energy (eV) |
|---|---|---|
| 0 | $3.02 \times 10^{-5}$ | 0.42 |
| **0.05** | $\mathbf{3.50 \times 10^{-4}}$ | **0.29** |
| 0.10 | $8.68 \times 10^{-5}$ | 0.32 |
| 0.15 | $8.05 \times 10^{-5}$ | 0.34 |
| 0.20 | $5.84 \times 10^{-5}$ | 0.37 |

### *3.5.DC Conductivity Study*

To confirm the ionic conduction in all the synthesized ceramic samples of $Li_{6.75+x}La_{3-x}Sr_xZr_{1.75}Ta_{0.25}O_{12}$ (x = 0 to 0.20) series, the DC polarization technique was used. Fig. 8 shows the DC conductivity graph of 0.05 Sr ceramic sample. The ionic transport number was calculated using the equation$t_i = (\sigma_{total} - \sigma_e)/\sigma_{total}$. Here, the silver paste was applied on both the faces of pellets and constant voltage of 1 V was applied for measuring the current through the sample. After some time, the current through the sample was almost constant and it is believed to be by electrons only [1,2,20]. Moreover, the calculated ionic transport number for 0.05 Sr ceramic sample is found to be $\geq 0.999$ which confirmed the predominant ionic conduction and negligible electronic contribution within the ceramic sample.

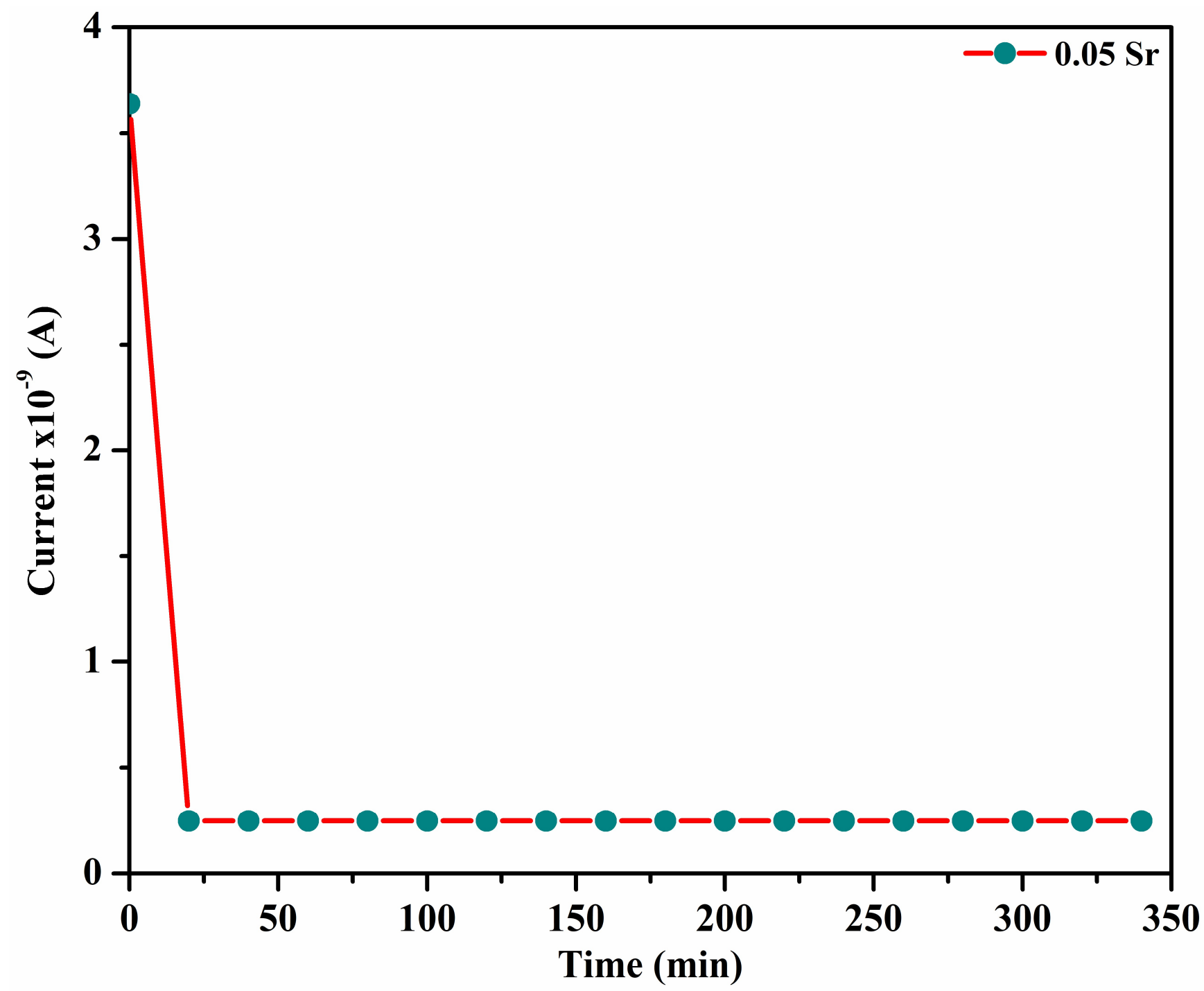

**Fig.8: DC conductivity graph of 0.05 Sr ceramic sample.**

## 4. Conclusions

The conventional solid state reaction method was employed to synthesize the series $Li_{6.75+x}La_{3-x}Sr_xZr_{1.75}Ta_{0.25}O_{12}$, where Sr (x) has been varied from 0 to 0.20 a.p.f.u.. All the prepared ceramic samples possessed the highly conducting cubic phase which was confirmed by X ray diffraction. The highest relative density with the dense surface morphology was obtained for 0.05 Sr ceramic sample. This can be due to the insertion of Sr along with Ta in LLZO, which acted as a sintering aid and helped in the densification of the crystal structure. The uniform distribution of all the elements in 0.05 Sr ceramic sample has been verified from its elemental mapping. The dense microstructure helped in the migration of Li ions, resulting in the highest room temperature ionic conductivity of 3.5 x $10^{-4}$ S/Cm for 0.05 Sr ceramic sample. Moreover, the minimum activation energy of 0.29 eV possessed by the 0.05 Sr ceramic sample showed that the optimum content of Sr not only channelized the Li ion migration pathways but also reduced the possible lattice distortion of the crystal. The predominant ionic conduction in 0.05 Sr ceramic sample was confirmed from the DC conductivity measurement. Thus, the 0.05 Sr ceramic sample can be a suitable choice as a solid electrolyte for ASSLIB's.

**Acknowledgement**

Authors like to acknowledge Department of Physics, VNIT, Nagpur for providing XRD facility governed by DST FIST project number SR/FST/PSI/2017/5(C). Authors also wish to acknowledge IIT, Bombay for providing the Impedance measurement facility.

**Funding**

This research did not receive any specific grant from funding agencies in the public, commercial, or not-for-profit sector.